# A Novel Design of a Parallel Machine Learnt Generational Garbage Collector

Vasanthakumar .S,

thisisvasanths@gmail.com
Anna University, Chennai, India.

## *ABSTRACT*

*The Generational Garbage collection involves organizing the heap into different divisions of memory space in-order to filter long-lived objects from short-lived objects through moving the surviving object of each generation's GC cycle to another memory space, updating its age and reclaiming space from the dead ones. The problem in this method is that, the longer an object is alive during its initial generations, the longer the garbage collector will have to deal with it by checking for its reachability from the root and promoting it to other space divisions, where as the ultimate goal of the GC is to reclaim memory from unreachable objects at a minimal time possible. This paper is a proposal of a method where the lifetime of every object getting into the heap will be predicted and will be placed in heap accordingly for the garbage collector to deal more with reclaiming space from dead object and less in promoting the live ones to the higher level.*

## 1. INTRODUCTION

The currently well-established generational garbage collector works in a way to filter the long-lived objects in the heap from the short-lived ones, since 80-98% of the newly allocated objects will be dead within a few million instructions or before they meet their first GC cycle [2]. The Generational GC attains this by dividing the heap into different regions of memory spaces called generations. All newly created objects will be allocated with space in the generational space division meant for the new objects, which is comparatively very small to the generational space division for the old objects, in order to have a Young generation GC which will be faster and frequent. As the GC cycles occur, whenever the space in this young generation is filled up, the surviving objects from this GC will be promoted to another space division meant for the surviving objects. This space acts like a buffer space where the objects are aged by swapping live objects between sub-divisions and clearing the dean objects through each GC cycle. Ones the objects have attained a certain age they are moved from here to the space division meant for the old objects. By this way the heap will have a group of filtered objects which have survived to the old space and will be expected to age here. This space for Old objects is bigger than the other two and the garbage collection will take place here less frequently and takes more time comparatively.

The problem with this method is that the garbage collections here are focused on filtering out the longer living objects from the short-lived ones and so ends up spending considerable amount of time in dealing with the live objects. The ultimate goal of the garbage collection is to reclaim the space allocated to objects which are dead to get the space ready for newer objects. So every encounter with a live object is waste of computational time since these activities are carried out as stop-the-world actions which means, all the current program threads will be paused while the garbage collection is done. As the number of program threads increases, the delay due to garbage collection also increases, and the time spent in checking reachability for live objects and promoting them can turn to be a counter-productive activity.

This scenario of entities being created, aged and reclaimed when dead, caries the computational advantages of a quintessential subject for the heuristic prediction over the lifetime of entities in a particular domain [1, 39]. This paper is a proposal of a method in which the expected life time of the objects getting created in the heap can be calculated in parallel to the application and sorted in the live data structure from the root, so that when a garbage collection cycle is initiated, the GC will encounter only with dead objects almost all the time and comparatively a very less number of live objects. The objects which the GC would not encounter will be promoted in parallel, to generation spaces where they are highly likely to die. By this way the pause times can be greatly reduced at situations where the application usually has a large number of long-living objects in its heap.

## 2. ASSUMPTIONS

Many different languages use automatic garbage collection as an integrated part of them. The Generational Garbage collection method has been implemented as a part of the JVM the .Net Framework and as part of many other languages as well. For feasibility of explanations, in this paper we will consider a java application running in a Sun JDK to be our subject. All our terminologies, and implementation details shall be in accordance to the JVM's Generation Garbage collection and JVM's Heap related for the rest of the paper. We will be using a sample java application named java2demo.jar as our subject to discuss the feasibility of prediction as the control flows for this application is very limited. Let us assume to have a JVM whose Garbage Collector code has been modified to inject a data retrieval module, which can log the details regarding objects creation, space allocation, current generation of an object, and reclamation. This particular injected module can be turn on or off when required by specifying the same as JVM arguments.

## 3. SOLUTION PROPOSAL

There are three important phases in the solution proposed through this paper. They are, Prediction of object life time in heap, structuring of data to hold dead objects first and reclamation of space parallel to promotion of objects.

### 3.1 Prediction

Computer programs are highly structured set of instructions and so a fully functional software application would have only a few finite different flows of execution. In the case of Object Oriented Programming, these control flows will determine which objects will be instantiated and which object's scope will be exited. These characteristics of software applications written in object oriented programming languages makes their objects creation pattern to be predictable in a given flow of execution. In this proposed method, the important module happens to be the prediction module. This a module which will run in parallel to the application threads and collects objects creation detail for a few initial test runs by injecting a data set retrieval code into the application's JVM's garbage collector module. This phase of running the application with injector codes will be purely testing intended and so the application's performance will be impacted due to the data collection retrieval for every object creation, promotion and reclamation. Once we have the test set in hand, the machine learning module will go through the test set and generates prediction data for each object, which will be the expected life time for the object in heap.

## 3.2 Structuring Reachability Data

The traditional approach by which the garbage collector in JVM will differentiate the dead objects from the lives one is based on their reachability form the root. Thus for each object, the GC will be traversing from the root until it either encounters the object or reaches the point where further traversal is not possible. According to this proposed model, the Root will hold only two children. One child will serve as the root for all the objects which are highly like to die before they meet their first GC cycle and the other child will be serving as the root for all the objects highly like to live longer or at the least long enough to survive their first GC encounter.

## 3.3 Parallel Reclamation and Promotion

Now, since we have two sub-roots under the main root, only one will be of the interest of the GC where objects residing are highly likely to be reclaimed of their space. Thus there will be one GC cycle running in parallel with another thread, which we shall call as the Object Promoter (OP). The OP will run through every child from its root where all the objects are expected to be alive. Here the OP will check for the expected life time of each object and will promote them accordingly to the respective generation space. If the OP encounters a dead object (which will occur often during the initial test runs to obtain the data set), the OP will act like a GC and reclaim its space.

# 4. DATA COLLECTION AND ANALYSIS

In order to verify how feasible the prediction phase of this model can be implemented, we can use the java2demo.jar application, as our subject and collect data during its runtime regarding its objects lifetime and analyze it. Using the jmap tool which is an integrated part of Sun JDK we can retrieve the histogram of the heap at any point during the runtime of the application. Further using the JvisualVM tool which also an integral part of the JDk with its plugin visualGC, we can have a graphical depiction of the GC cycles and the objects promotion.

The jmap's –histo and –histo:live options can be used to get the list of class names, number of instances for each class name and their size in bytes for all the objects and only the live objects respectively. Using the jvisualVM we can learn when the GC cycle is taking place. A –histo from jmap after each GC cycle, followed by a –histo:live can give us enough data to understand that the number of live instances of a particular class after a particular GC cycle happens to be almost the same every other time we run the application with minor variation which is due to the interruptions from our jmap commands and jvisualvm.

The difference between two –histo:live lists taken in consecutive GC , (i.e.) removing the old list's number of instances from the new list's number of instances will give us the number of objects surviving the Eden Space and their class name and size. This data will also be consistent for every other run of the application. As an example, running the jmap –histo after the first GC cycle gave me a set of results, a very few of them for comparison are as follows,

| num | #instances | #bytes | class name |
| --- | --- | --- | --- |
| 4: | 13203 | 1511320 | <constMethodKlass> |
| 27: | 15736 | 251776 | com.sun.media.sound.ModelSource |
| 137: | 659 | 10544 | javax.swing.ArrayTable |
| 891: | 3 | 72 | java.awt.Polygon |
| 1305: | 1 | 24 | sun.nio.cs.US_ASCII |

Fig. 1 Jmap –histo Output 1

And running jmap-histo after the first GC cycle during the second run of the application gave me the another set of results and the data for the same five classes are as follows,

| num | #instances | #bytes | class name |
| --- | --- | --- | --- |
| 4: | 13199 | 1510976 | <constMethodKlass> |
| 30: | 15736 | 251776 | com.sun.media.sound.ModelSource |
| 163: | 659 | 10544 | javax.swing.ArrayTable |
| 911: | 3 | 72 | java.awt.Polygon |
| 1223: | 1 | 24 | sun.nio.cs.US_ASCII |

Fig. 2 Jmap –histo Output 2

Now from the above data we can infer that the number of objects that are being formed, and number of objects surviving the Eden space will be the same for every run of the application at a given instance. Here, it happens to have a slight variation in the number of instances for classes which has huge number of instances and that is due to our interruptions, as mentioned earlier. On the other hand, the similarity is also mainly because the application has only minimal number of control flows in which it can progress. Hence the prediction of objects life time over these applications can be simple. Further for applications which are widely dynamic in their flow of control, for example an application that reacts in a different way to each different input from the user, the prediction can get tougher but still possible. The only difference will be the effect of presence of one object over the other will have to be taken into consideration. By which the presence of a given number of different instances can be used to predict the control flow of the application and then the creation and deletion of objects in that control flow. Such a scenario where one feature of an entity has a direct influence on another feature of the same entity, or another entity in the same domain, the prediction of data can be performed through probabilistic inference [4]. This can be achieved by using widely implemented conditional probability theorems such as the Bayesian Network discussed in the next section.

# 5. BAYESIAN NETWORK

Bayes' Theorem is a theorem of probability theory which can be seen as a way of understanding how, the probability that a theory is true; is affected by a given piece of evidence. It has been used in a wide variety of contexts, ranging from marine biology to the development of "Bayesian" spam blockers for email systems. The Bayesian network will be the right approach to this scenario, since our prediction in here is based on conditional probability (i.e.) the probability of event B to occur given that the an event A has occurred [5]. Here the event B refers to the survival of an object for a specific time, given that a number objects from another class already exist or the number of GC cycles spent or the same object has survived the Eden space etc.

## 5.1 Features and Probability

Bayes' theorem expresses the conditional probability, or 'posterior probability', of an event A after B is observed in terms of the 'prior probability' of A, prior probability of B, and the conditional probability of B given A, denoted B |A. Bayes' theorem is valid in all common interpretations of probability.[6]

Bayes' theorem provides an expression for the conditional probability of A given B, which is

$$\Pr(A|B) = \frac{\Pr(B|A)\Pr(A)}{\Pr(B)}$$

The features based on which the probability has to be calculated will depend on the feature data that can be retrieved from the heap during the run time of the application. As per our earlier assumption, let us assume that there are 'n' different features which can be recorded or calculated, during or after the runtime of the application from the data our injected code can retrieve during the initial few test runs. The features set will include features like the object's class name, depth of hierarchy from the root parent, number of same class's objects formed, number of same class's objects surviving the Eden space, size of the object, etc. For each of these features, the Bayes theorem will be used to calculate the posterior probability of the particular object to survive the generation is currently in.

$$\Pr(Y|Fn) = \frac{\Pr(Fn|Y)\Pr(Y)}{\Pr(Fn)}$$

Fig. 4 Bayes Theorem Implementation

Here, Pr(Y|Fn) is the probability of an object to survive given that a feature meets a condition. Pr(Fn|Y) is the probability of the feature to meet the condition given that the object has survived.

Pr(Y) is the probability of the object to survive the current generational space and Pr(Fn) is the probability of the feature to meet the conditions.

For example, taking each of the feature into consideration such as,

F1 = depth of hierarchy
F2 = className
.
.
Fn = size

The probability of the object to survive this generation given that the feature Fx holds this particular value can be calculated.

For features that hold static value such as the class name and depth of hierarchy, the probability can be calculated just once, whereas for the dynamic ones such as the number of instances will have to be calculated for every GC cycle.

$$P(X1, \ldots, Xn) = \prod_{i=1}^{n} P(Xi \mid \pi(Xi))$$

Here, $\pi(Xi)$ stands for the set of parents (direct ancestors) of Xi.

By constructing a Bayesian Network as proposed above, we will arrive at a Directed Acyclic Graph where nodes are variables and edges indicate casual influences. A Bayesian network implicitly defines a joint distribution.

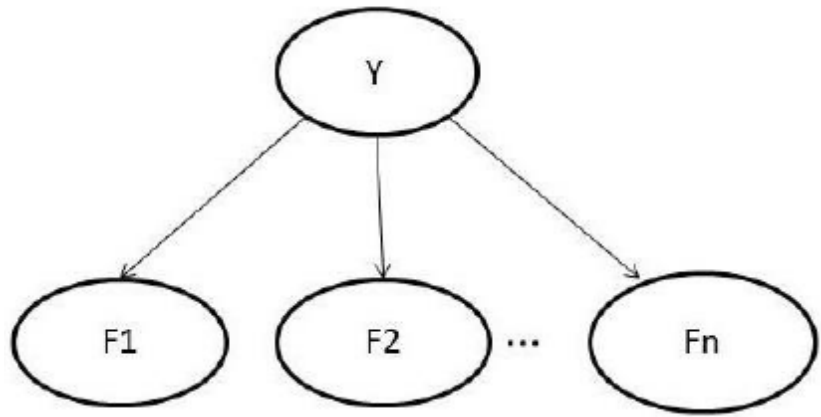

Fig. 6 Bayesian Network

This joint distribution will on the whole calculate the probability of an object to survive a particular generational space. On the whole for our implementation we will need to calculate this for an object only twice, i.e. the probability that an object will survive the Eden space, and the probability of an object to reach the tenured space.

### 5.2 Threshold and Decision making

Since the Bayesian network will give us a probability of an object to survive a particular generation space or reach a particular generation space, the output will be a value between 0 and 1, inclusive. Now, we will have to find a threshold value which will be the deciding factor above which the object will be surviving or moving to a space, below which the object won't. This value can be a tunable factor which can be set, based on the performance of the algorithm over the application for improved results. But we need to consider the fact that a false prediction of an object to not to survive a generational space can be tolerated as the GC will anyway promote the object to the next generation of buffer space, But a false prediction of an object to survive a generation space cannot be tolerated as an 'about to be dead' object will be promoted which will cause further computational expenses. And so the threshold for this scenario will have to be generally high, for instance a threshold value of 0.8 will perform better than the threshold 0.55.

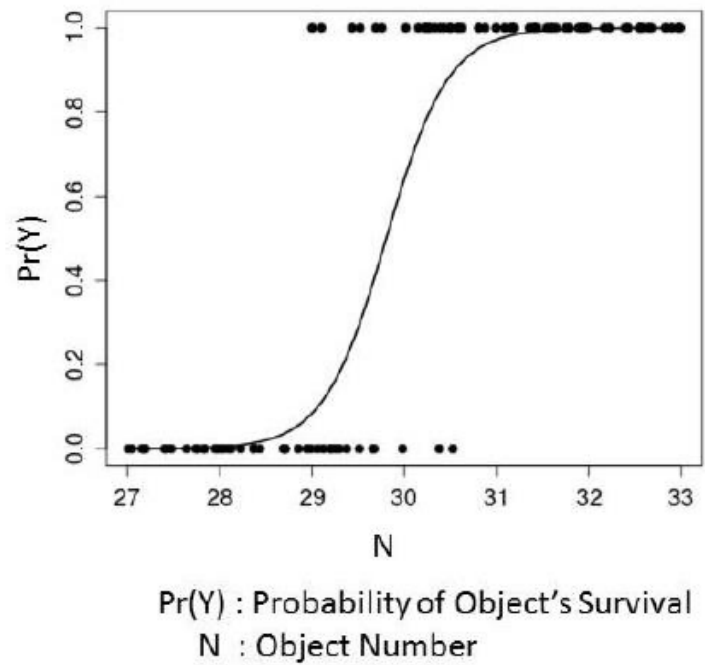

Pr(Y) : Probability of Object's Survival
N    : Object Number

Fig. 7 Logistic Regression

## 6. PRECISION AND RECALL TRADE OFF

This prediction method proposed for the scenario has two cases of false predictions. One is a False Positive, which is that the algorithm predicts that the object in question will survive the space but it would turn out to die due to which we would promote a dead object. The second case is a False Negative where the algorithm predicts the object to die but instead survives the GC cycle, in which case the GC will promote the object. Considering both the cases we can clearly decide that a False Negative is a tolerable scenario where as the False Positive cannot be tolerated. Thus we will have to tune the prediction module to have absolutely no False Positive which might give space to a few false negative. Such a prediction system is High Precision Classifier system.

$$Precision = \frac{True\ Positives}{No.\ of\ Predicted\ Positives}$$

$$Recall = \frac{True\ Positives}{No.\ of\ Actual\ Positives}$$

Fig. 8 Precision and Recall

This will be a tradeoff between the precision and recall of the classifier where we will choose to have higher Precision which will result in less recall. In order to set a classifier system to have higher precision we will have to assign a higher value to the threshold. For example, setting up the threshold to have a value of 0.8, the system will be a high precision classifier. By which the prediction module's confidence will be higher for a Positive prediction (i.e.) an object will survive the current GC cycle. This confidence will increase the reliability over the prediction module to predict objects survival.

## 7. DATA STRUCTURING

This method of garbage collection with predicted life time of objects will need to have a slightly different implementation of the data structure used for to detect the dead objects from the live one. The traditional Generational GC will do this by checking the reachability of an object from the root. But in this case we will maintain a root node which has two child nodes, each one acting as a root for a map. One child will act as the root for objects which are highly likely to die in this current GC cycle, whereas the other child will act as the root for the objects which are highly likely to survive the current GC cycle. The child holding the objects which are about to die will be acting as the root for the GC , whereas the other child holding the objects about to be promoted will act as the root for the OP. This way, the GC will use a root where a very small set of objects are reachable, which are the ones predicted to die in the GC cycle and in turn survived (False Negatives). The OP will deal with the root from where almost all the reachable objects are. This will result in a big cut down of the GC cycle pauses which runs as a 'stop-the-world' process.

Once an object is allocated space in the Eden generational space, the object will be under the OP's root. The OP will go through each object from the root and predict the life time of the objects. As the OP moves through the tree every object whose probability to survive the upcoming GC cycle is low will be made available from the GC root and removed reference from the OP root. Every dead object that the OP faces will be reclaimed of memory and every object which is likely to survive the GC cycle will be marked with its expected life time. The objects will also be sorted in a way that the ones with higher life time expectancy will be closer to the root than the ones which have less life time expectancy. During the GC cycle , the OP will run through the reachable objects from the OP root and promote the objects to their respective generational spaces where they are likely to die. Things under the GC root will be the same process as the traditional Generational Garbage Collection process.

## 8. PARALLELIZATION OF PROCESSES

This Garbage Collection method has been designed to be suitable for implementation on multiple processor machines. For each application for which we plan to use this Predictive GC, we need to get a number of initial trial runs to obtain the data set to work on which number will be based on the complexity of the application. Once the data set has been acquired and processed, apart from the heavily reduced GC pause times, every other action in this proposed method of GC can be parallelized and performed with-out disturbing the run of the application threads. The OP can run in parallel with the GC since they don't share the same root and can promote the objects to the expected generational spaces where they are highly likely to be reclaimed of their spaces. The OP can also predict the life time of objects in parallel to the application threads.

## 9. CONCLUSION

I have proposed the description of a Parallelized Machine Learnt Generational Garbage Collector which uses Bayesian Network to predict and manage the objects in the heap accordingly to reduce the time spent by the GC in dealing with live objects. This proposed model, when implemented is highly likely to result in reducing the work load of the GC in each generational phase as it will not deal with the live objects if the prediction is 100% accurate. The pause times will be greatly reduced in applications containing a large amount of live objects in their Eden and Survivor spaces. Since the Generational Garbage Collection is used along with many programming languages which are being used widely across the globe for application engineering, such a model can improve the performance of the GCs which in turn will reflect as a performance increment over the application. This model is based on the assumption that the application taken as a subject doesn't have much different control flows which affects the objects creation pattern. As a future enhancement to this paper, I will be carrying out a research to amend this model to be fit for predicting objects lifetime in a highly complex application with a large number of different control flows.

# REFERENCES

[1] Henry G. Baker., Infant mortality and generational garbage collection in SIGPLAN Notices 28(4), April 1993, pages 55–57.

[2] Java SE 6 HotSpot[tm] Virtual Machine Garbage Collection Tuning, in www.oracle.com

[3] Yama: A Scalable Generational Garbage Collector for Java in Multiprocessor Systems February 2006 (vol. 17 no. 2)

[4] Richard E. Neapolitan in Learning Bayesian Networks

[5] David Barker in Bayesian Reasoning and Machine Learning

[6] http://www.bayesian-inference.com/bayestheorem

[7] Tim Brecht, Eshrat Arjomandi, Chang Li, Hang Pham, Controlling garbage collection and heap growth to reduce the execution time of java applications in Proceedings of the OOPSLA'01 Conference on Object Oriented Programming Systems Languages and Applications, ACM Press (2001).

[8] Hans-Juergen Boehm, Alan J. Demers, Scott Shenker, Mostly parallel garbage collection in ACM SIGPLAN Notices, 26 (6) (1991).

[9] Hans-Juergen Boehm, Mark Weiser, Garbage collection in an uncooperative environment in Software—Practice and Experience, 18 (9) (1988).

[10] Zaman, W.U., Ahmad, S.A., Abbas, A., Qadeer, A., A novel design of a generational garbage collector in Students Conference, 2002. ISCON '02. Proceedings. IEEE (Volume:1 )

[11] H. Lieberman and C. Hewitt, "A real-time garbage collector based on the lifetime of objects," in Communications of the ACM 26, pp 419-429, June 1983.

[12] Sun Wenjing ; Dept. of Math., Xidian Univ., Xi'an, China ; Yang Youlong ; Li Yangying, Learning Bayesian Network Classifier Based on Dependency Analysis and Hypothesis Testing in Intelligent Human-Machine Systems and Cybernetics (IHMSC), 2013 5th International Conference on (Volume:1)